 \documentstyle[11pt]{article}
\def\ll{\label}
\def\re{\ref}
\def\c{\cite}

\def\r1{(\ref{$1})}

\def\ti{\tilde}

\def\ba{\begin{array}{c}}

\def\ea{\end{array}}

\def\si{\sigma}

\def\ov{\over}
\def\ha{{1\over 2}}

\def\l{\left}
\def\l({\left(}
\def\r){\right)}
\def\r{\right}

\def\la{\lambda}
\def\al{\alpha}

\def\be{\begin{equation}}
\def\bc{\begin{center}}
\def\ec{\end{center}}
\def\bit{\begin{itemize}}
\def\eit{\end{itemize}}
\def\ee{\end{equation}}
\def\ed{\end{document}}
\def\bea{\begin{eqnarray}}
\def\eea{\end{eqnarray}}
\def\efr{\end{flushright}}

\begin{document}
\title{Ultralocal  solutions for  quantum  integrable nonultralocal  models 
} 
\vskip 1cm

\author{
Anjan Kundu \footnote {Email: anjan@theory.saha.ernet.in, \ Phone:
  +91-33-337-5345(-49); Fax: +91-33-337 4637
} \\  
  Saha Institute of Nuclear Physics,  
 Theory Group \\
 1/AF Bidhan Nagar, Calcutta 700 064, India.
 }
\maketitle
\vskip 4 cm

\begin{abstract} 
 A challenge in the theory of integrable systems is to show for every
 nonultralocal quantum integrable model, a possible 
connection to an ultralocal model. Some of such gauge
 connections were discovered
earlier. We complete the task by identifying  the same for the remaining
ones along with two  new models. We also  unveil the underlying
algebraic structure for  these nonultralocal models. 
\end{abstract}

\vskip 2.cm

 {\it PACS numbers}: 
 02.30.Ik,
 03.65.Fd,
  02.20.Uw,
11.10.Lm

\vskip 1.cm

{\it Key Words}:  Nonultralocal quantum integrable models, Braided Yang-Baxter
 equation, Discrete current algebra, Gauge connection with ultralocal models

\newpage
\noindent 1. {\bf Introduction}

 Impressive success has been  achieved  
in the theory of quantum
  integrable systems during the last twenty years. However  
the major progress made   and the important  results obtained in the subject
 are limited mostly to a class of models, known as the 
ultralocal (UL) class. The Liouville model,  sine-Gordon (SG) 
model in laboratory coordinates, relativistic Toda chain, 
 nonlinear Schr\"odinger equation (NLS) etc.
are few   well known examples of this class. 
 The UL models  must satisfy the ultralocality condition, i.e.   
  their representative Lax operators 
 at different lattice
 sites  must commute:
$ L^{ul}_{2k}(\mu)L^{ul}_{1j}(\la) =L^{ul}_{1j}(\la)L^{ul}_{2k}(\mu) \ \ 
\mbox {for } j \neq k,$ 
which is essential for extending the Yang-Baxter equation (YBE):
$ ~{R}_{12}(\la-\mu)L^{ul}_{1j}(\la)L^{ul}_{2j}(\mu)
= L_{2j}^{ul}(\mu) L_{1j}^{ul}(\la){R}_{12}(\la-\mu) ~$
 to its
global  form: 
$ ~{R}_{12}(\la-\mu)T^{ul}_{1}(\la)T^{ul}_{2}(\mu)
= T_{2}^{ul}(\mu) T_{1}^{ul}(\la){R}_{12}(\la-\mu) ~$,  for the monodromy
matrix $T_{a}^{ul}(\la)=\prod_{j=1}^N L_{aj}^{ul}(\la)$. Note that 
$a=1,2$ labels here the auxiliary or matrix  spaces, while $j=1,2, \ldots, N$
denotes the associated quantum spaces.  For the models with
periodic boundary condition  the global YBE leads in turn  to the required 
trace identity $[\tau(\la),\tau(\mu)]=0$,  for the transfer matrix $\tau(\la)=
tr_a(T_{a}^{ul}(\la))$,  
  establishing
 the quantum integrability of the system by showing 
 mutual commutativity:
$[c_n,c_m]=0$, for
the set of conserved operators
obtained  
 from $ \tau(\la)=\sum_n c_n\la^n$ or from any other function of
it. 
The eigenvalue problem for all these conserved operators including the
Hamiltonian  of such models 
can be solved exactly by the powerful algebraic Bethe ansatz, which
effectively exploits the global form of YBE \c{Faddeev}.   
 The ultralocality condition
  reflects  generally   the fact that   the basic fields
involved in the Lax operator of an UL model are of  canonical nature, which 
commute trivially   at space-like separated  points $x \neq y.$

 However,  there
exists   on the other hand another rich class of    integrable 
 models  which  violate  the ultralocality condition   
and  make their  quantum description 
  through standard   YBE formulation difficult.  In such nonultralocal (NUL)
 models the basic fields are either of noncanonical
type having nontrivial commutators at $x \neq y$ or their associated Lax
operators contain space-derivatives of canonical fields. Nevertheless
  there is now a considerable 
number of NUL systems, namely, 
nonabelian   
 Toda chain \c{korepin}, the quantum mapping \c{Nijhof},   
  model related to the  Coulomb gas picture of CFT  \c{babelon}, 
current  in WZWN  \c{alekfad}, 
integrable  model on  moduli space   
\c{alex} and the quantum  mKdV 
\c{mkdv95,mkdv02},
 for which the  
  quantum integrability  is established 
 and the
   braided extensions of  
YBE  have been  formulated \c{Maillet91,hlav94,hlavkun96}. We shall call
such NUL models genuine quantum integrable  models. However, there are some
NUL models like the SG in light-cone coordinates and the
 nonlinear $\sigma$-model,
for which the 
 quantum integrability through braided YBE still
remains  unsolved.  

Nevertheless, for the complete understanding of NUL models
and for fully  exploiting the powerful 
 machinery of quantum integrable systems
 developed already for  UL models, it is highly desirable 
 to find a possible connection for these NUL systems to  UL models. 
Though in general   such a connection is not guaranteed, 
   an UL solution was first 
discovered    for the nonabelian 
 Toda chain   using a operator dependent local gauge transformation
 \c{korepin}.
 Subsequently, similar relation was unveiled also
for other quantum NUL systems like current algebra in WZWN model \c{alekfad},
 Chern-Simon theory related integrable  model on  moduli space   
\c{alex} and
 the  mKdV model
\c{gmkdv01}.

 It is therefore tempting to conjecture that such a connection
to  UL model should exist for every NUL system, at least for the 
genuine quantum integrable  models that respect the braided YBE.
 However the problem is that, the well known NUL
  models like  the Coulomb gas picture of CFT (CGCFT)
 \c{babelon} and the quantum mapping  \c{Nijhof},  though 
fit well into the  braided YBE formulation
 \c{hlavkun96} are
 not known to have any established UL connection.
Moreover for the quantum light-cone SG (LCSG) model, 
 not only its explicit UL relation   
is unknown, but also, as  mentioned already,   its 
   quantum integrability  through braided YBE could not be   formulated.
Our aim here is to find  the required  NUL-UL connection for all these 
   models and thus to resolve the eminent
 problem  for the validity of the
 above
conjecture, at least for the genuine quantum integrable
   models discovered till today.
Along with the above result we  find also a new  NUL  sine-Gordon (NSG)
 like system
and able to  identify  its connection with a  Liouville-like UL
model on a lattice. A preliminary announcement on  two new models investigated
here, e.g. LCSG and NSG, has been  made in \c{kun02}.
\\ \\ 
\noindent 2. {\bf Nonultralocal  models and their ultralocal connections}

The NUL models of our concern satisfy the
braiding relation
\begin{equation}
 L^{nul}_{2 j+1}(\mu)Z_{12}^{-1}L^{nul}_{1 j}(\la)
=L^{nul}_{1 j}(\la) L^{nul}_{2 j+1}(\mu)
\ll{zlzl1u}\end{equation} 
involving nearest-neighboring (NN) sites,
together with the braided YBE (BYBE) \c{hlavkun96}
\begin{equation}
{R}_{12}(\la-\mu)L^{nul}_{1j}(\la)L^{nul}_{2j}(\mu)Z_{12}^{-1}
= L_{2j}^{nul}(\mu) L_{1j}^{nul}(\la)Z_{21}^{-1}{R}_{12}(\la-\mu).
\ll{bqybel}\end{equation}
Note that along with the $R$-matrix as in UL models an additional braiding
matrix $Z$ enters into these equations.
Remarkably  the nonultralocality (\re{zlzl1u}) can  be incorporated
in the  BYBE to obtain its global form for the corresponding monodromy
matrix  $T_{a}^{nul}(\la)= L_{aN}^{nul}(\la) \cdots L_{a1}^{nul}(\la)$ with
$a=1,2$  as
\be {R}_{12}(\la-\mu)T^{nul}_{1}(\la)Z_{21}^{-1}T^{nul}_{2}(\mu)Z_{12}^{-1}
= T_{2}^{nul}(\mu)Z_{12}^{-1} T_{1}^{nul}(\la)Z_{21}^{-1}{R}_{12}(\la-\mu),
\ll{bqybet}\ee 
which leads again to  the trace identity establishing the quantum
integrability for the NUL models.
 It may be noticed that
unlike the YBE
the global  form of   BYBE  differs a bit  from its local  form. This happens 
because
  for the periodic models with $N+1 \equiv 1$,
 the Lax operators $L_{aN}^{nul}(\la)$
 and $ L_{b1}^{nul}(\la)$, $a,b=1,2$   become in fact  NN operators  exhibiting 
nontrivial
commutation relations  due to (\re{zlzl1u}).   
It can be   seen easily  that  by putting  $Z=1$ 
in the above  NUL equations, i.e.
   braiding relation  and the
 BYBEs, one  obtains  
 the corresponding equations, i.e. 
the ultralocality condition and the standard YBEs 
 for the UL models.

The  BYBE (\re{bqybel})  provides similar Hopf algebra structure 
  to the 
integrable NUL models, though  braiding relation 
(\re{zlzl1u}) induces now braided algebra property by modifying the
multiplication rule \c{hlav94,hlavkun96}.
At the same time  the trace identity for the  NUL models 
as obtained from (\re{bqybet}) is also modified   
to
$tr_{12} (T^{nul}_{1}(\la)Z_{21}^{-1}T^{nul}_{2}(\mu)Z_{12}^{-1})=
 tr_{12}(T_{2}^{nul}(\mu)Z_{12}^{-1} T_{1}^{nul}(\la)Z_{21}^{-1}) $
and consequently   its crucial factorizability depends heavily on the 
the structure of the braiding matrix $Z$,  adding another difficulty in
tackling such models \c{hlavkun96}.
 Therefore  though some progress has been made 
, as mentioned above, in direct analysis of
integrable NUL  models, identifying  explicit gauge relation of
such systems to UL models   would be significantly important for getting
insight into their underlying algebraic structures
 as well as their exact solutions.
Following the  idea of \c{korepin} we aim to   
 find such  UL source   models for genuine quantum integrable 
   NUL models  through a gauge transformation like  
 $L^{nul}_j =D_{j+1}L^{ul}_jD_{j}^{-1}$, such 
that  $L^{ul}_j$  satisfies the  YBE  along with the ultralocality.
 This in turn is guaranteed, as can be shown by some simple algebraic
manipulation, by
 the condition \be
D_{aj}L^{nul}_{bj}D_{aj}^{-1}=L^{nul}_{bj}Z_{ab}^{-1},\ \  a,b=1,2, \ 
 a \neq b, \ll{cond}\ee 
provided the gauge operator $D_j$ satisfies the YBE together with a related
equation as
\be
{R}_{12}(\la-\mu)D_{1j}D_{2j}
=D_{2j} D_{1j} {R}_{12}(\la-\mu), \ \  
{R}_{12}(\la-\mu)D_{2j}D_{1j}
=D_{1j} D_{2j} {R}_{12}(\la-\mu).
\ll{ybeD} \ee
However in the particular case when  $ [D_{1j}, D_{2j}]=0,  $ both the
equations (\re{ybeD}) reduce to a single condition 
$ [R_{12},\
D_{1j}D_{2j}]=0 $, which actually holds true for  gauge operators in all
the examples we have considered here.
Moreover when  $D_{bj} $ and $Z_{ab}$
 mutually commute (which is true  for all our examples except that with the 
quantum mapping), 
  condition
(\re{cond}) holds also for the gauge related  $L^{ul}_j $.
Interestingly the monodromy matrices for these periodic models are 
 linked  now  as 
\be
 T^{nul}=\prod_j^N L^{nul}_{j}=  D_{N+1}(\prod_j^N L^{ul}_{j} ) D_{1}^{-1}
= D_{1}T^{ul} D_{1}^{-1},
\ll{monrel}\ee
  and satisfy  the relation 
\be ~ 
D_{a1}T^{nul}_{b}D_{a1}^{-1}=T^{nul}_{b}Z_{ab}^{-1},\ \  a,b=1,2, \ a \neq b,
\ll{mon}\ee
  as can be shown starting from (\re{cond}) with the assumption 
$[D_{aj+1},L^{nul}_{bj}]=0, \ a \neq b,$, which is fulfilled in  our cases.
We should mention here that though the global BYBE (\re{bqybet}) can be derived 
from its local form and the braiding relation independent of its conjectured
relation with the UL model, by using such   relations (\re{monrel})  and
 (\re{mon}) one can
actually derive 
(\re{bqybet}) starting  from the standard YBE for the corresponding $T^{ul}$. 
 Inspired by the fact that the UL model must have
the same $R$-matrix as its NUL counterpart related by a gauge transformation,
 we intend to  seek our $L^{ul} $ solutions
 using  the  generating scheme of   \c{kunprl} based on the $R$-matrix
classification.

  Recall that  the
 well known  trigonometric  $4\times 4$ 
$R(\la)$-matrix is  defined by its  
nontrivial 
  elements 
\be \ \ R^{11}_{11} = R^{22}_{22}= \sin  (\la+\al),
\  R^{12}_{12} = R^{21}_{21}= \sin  \la, \ R^{12}_{21} = R^{21}_{12}=
 \sin \al ,
\ll{R-mat} \ee  
while the   ancestor Lax operator of  quantum  UL  models  
  associated with (\re{R-mat}) and satisfying the YBE 
 may be  given \c{kunprl}   by
\be
L^{t-anc (ul)}_k{(\la)} = \left( \begin{array}{c}
  {c_1^+}{\xi} q^{ S_k^3}+ {c_1^-} {1 \ov \xi} q^{- S_k^3}\qquad \ \ 
   S_k^-   \\
    \quad  
    S^+_k    \qquad \ \  {c_2^+}{\xi} q^{- S^3_k}+ 
{c_2^-}{1 \ov \xi} q^{S^3_k}
          \end{array}   \right), \quad
         q=
e^{i \alpha } , \ \ {\xi}= e^{i \la } \ll{L} \ee
or multiplying it by $\sigma^i, i=1,2,3$ due to a symmetry of 
(\re{R-mat}) as $[R, \sigma^i\otimes \sigma^i]=0$.
The  basic  operators in (\re{L}) 
  satisfy a   quadratic quantum algebra
\be
 [ S_k^ {+}, S_l^{-} ] =
  4 \delta_{kl}  \sin \al( M^+{\sin (\al 2  S^3) } + {M^- } 
{\cos (\al 2  S_k^3 )} ) , \ \ \ \ 
 [S_k^3,S_l^{\pm}] = \pm \delta_{kl} S_k^{\pm},  \quad  [M^\pm, \cdot]=0,
\ll{Alg} \ee
which is a Hopf algebra with well defined coproduct and with 
$c^\pm_a, a=1,2$ and 
  $ M^\pm=\pm \ha  \sqrt {\pm 1} ( c^+_1c^-_2 \pm
c^-_1c^+_2 ) $ 
being  sets of   central elements.
 Reductions of the general $L$-operator (\re {L}) (modulo its product with
$\sigma^i$), as shown in   \c{kunprl}, 
generate  in a systematic way the Lax operators of
  integrable UL models  having the
same $R$-matrix (\re{R-mat}).
 At the undeformed $q \to 1$ limit the whole
procedure is repeated for the rational $R$-matrix.

We  focus now on the NUL models under  consideration
for explicit  solution of
  their  UL  sources. Only 
 the simplest $SU(2)$ case is considered here;   
 more general group admissible to some models
 can be treated  analogously.
\\
2.1. {\it Coulomb gas picture of CFT} 

The NUL nature of this CGCFT model \c{babelon} follows from the   
 Drinfeld-Sokolov  problem $
\partial_xQ=L(x)Q$,
since the linear operator \be L(x)
= v(x) \si^3 + \si^+ \ll{coul-lax}\ee 
 contains  a  current-like  field 
$\{v(x),v(y)\}=-i \al \delta ^{'}(x-y).$
Quantum and discrete version of  this linear operator: $L_k$
was   shown to  
 satisfy the spectral parameter-less analogs of braided equations 
(\re{zlzl1u})
and (\re{bqybel}) with the braiding matrix $Z$ as presented below
and the limit of  (\re{R-mat}): 
$  R(\la \to \infty) \to R^+_q $ acting 
as the $R$-matrix.

Note that 
by  fixing
 the central  elements in (\re{L})
as
 $c^+_1=\Delta$, with $\Delta$  being  the lattice constant 
and all the rest $c$'s as zeroes, the algebra (\re{Alg}) reduces simply to
\be   [ S_k^ {+}, S_l^{-} ] =
0  \ \ \ \ [S_k^3,S_l^{\pm}] = \pm \delta_{kl} S_k^{\pm} \ll{tcalg}\ee
 and   we may find a realization for the
generators as \ \be 
S_k^ {\mp}=e^{\pm i p_k},  \ S_k^3 =   u_k-{1 \ov \al} \ p_k  \ll {cgcft}\ee
reducing the 
 Lax operator  (\re{L}) (after multiplying it from right by 
 $ \sigma ^1 $) to 
\be
L^{ul}_k(\xi)=e^{i p_k\si ^3} + \Delta \xi e^{i(\al u_k-p_k)} \si^+,
\ll{coul-ul}\ee
 in canonical
variables $[u_k,p_j]= i \delta_{kj}$. We may conclude that (\re{coul-ul})
 represents  a relativistic Toda chain (RTC) 
like  integrable UL model \c{kunrtc} and 
   satisfies the YBE with the 
$R$-matrix (\re{R-mat}), as a consequence of the general argument of 
\c{kunprl}.
    We observe that a 
local gauge matrix $D^{(1)}_j=e^{-{i \ov 2}  \al  \si ^3 u_j } $
along with a  change of  basic operators: 
 \be ~ v^+_k=-{\al \ov 2}(u_{k+1}-u_{k}) -p_k, \ \ 
v^-_k={\al \ov 2}(u_{k+1}-u_{k}) -p_k 
\ll{map} \ee
can transform 
(\re{coul-ul}) to 
\be L^{nul}_k(\xi)=e^{-i v^-_k\si ^3} + \Delta \xi e^{i v^+_k} \si^+ ,
\ll{coulomb}\ee
 NUL properties of which are induced by the discrete version of the 
current algebra  for $v^\pm_k$:
\be
[v^\pm_k, v^\pm_l] = \pm i {\al \ov 2} (\delta_{k,l+1}- \delta_{k+1,l}),
\ \ 
[v^+_k, v^-_l] =i {\al \ov 2} (\delta_{k+1,l}-
2\delta_{k,l}+\delta_{k,l+1}).
\ll{cr12}\ee

One may check  that 
  (\re{coul-ul}) as well as 
 (\re{coulomb}) satisfy   
(\re{cond}) with the  braiding matrix
   $Z^{(1)}=e^{- {i \ov 2} \al \si ^3 \otimes
\si ^3} $,
as  found in
  \c{babelon}. This guarantees in turn that 
the   NUL integrability relations
(\re{zlzl1u}), (\re{bqybel})
and (\re{bqybet}) must hold for the  Lax operator (\re{coulomb}). 
We observe further that 
 at the continuum limit $\Delta \to 0,
 $ $v^\pm_k \to\sqrt \pm 1 \Delta v(x)$ yield the current-like field
 and  consequently
   (\re{coulomb}) reproduces through $L^{nul}_k(\xi)= I +\Delta L(x,\xi)
+O(\Delta ^2)$, $L(x, \xi)
= v(x) \si^3 + \xi \si^+  $, i.e.   the field 
 Lax operator (\re {coul-lax})
 of \c{babelon}, though generalized here to  include  
 spectral parameter $\xi$ and satisfy
 the standard YBE, which shows also its novelty.
 Thus (\re{coul-ul}) is proved 
to be   the required UL source solution  for the
 NUL integrable  CGCFT model.
\\
\noindent 2.2. {\it NUL sine-Gordon like model}

For  introducing a new discrete  NUL  sine-Gordon (NSG) type
 model, we start from an UL  
  Lax operator 
\be L^{ul}_k{(\la)} =
   e^{i( \la \si ^3 -p_k)} + 
   e^{i(\al u_k-p_k)}\si ^+ +
    e^{-i(\al u_k+p_k)}\si ^- , \ \ \xi =e^{i \la} . \ll{llm-ul}\ee
Note that like the previous model it can again be derived  from the same    
general  Lax operator ({\re{L}), though now with a different parameter
choice: $c^-_1= c^+_2=0,\   c^+_1= c^-_2=1,$
 which   reduces the
  underlying general algebra (\re{Alg}) 
 to an interesting exponentially
deformed   algebra 
\be
 [ S_k^ {+}, S_l^{-} ] =
  -2i\delta_{kl} \ e^ {2 i\al S_k^3}   \sin \al, 
\ \ \ [S_k^3,S_l^{\pm}] = \pm \delta_{kl} S_k^{\pm}. 
\ll{eAlg} \ee

Note that  such reductions, which  
  must represent  
 integrable UL  models associated with the trigonometric $R$-matrix
(\re{R-mat}) fall in the class of discrete 
  Liouville model \c{liouville} and with a realization of (\re{eAlg}):
\be S_k^ {\mp}=e^{\pm i(\al u_k \mp p_k)},  \ S_k^3 =   -{1 \ov \al} p_k 
\ll{nsg}\ee
 we can derive (\re{llm-ul}) directly from (\re{L}).

Remarkably, in spite of the  different underlying algebra and its  
realization from  the previous model,  (\re{llm-ul})  
permits us to choose  the  gauge operator again  in the 
 same
form as the previous one:
$D^{(2)}_j\equiv D^{(1)}_j=e^{-{i \ov 2} \al \si ^3 u_j}. $  
  Performing therefore  this gauge transformation on (\re{llm-ul})  
and 
changing to the   current-like operators 
  $v^\pm_k$    using the map (\re{map})  we derive finally
    the  Lax operator of our new NSG model as 
\be
L^{nul}_k{(\la)} = \left( \begin{array}{c}
  \xi \  W_k^+\qquad \ \ 
   W^+_k  \\
    \quad  
    W^-_k \qquad \ \   
{1 \ov \xi} \ W^-_k
          \end{array}   \right), \quad
W^\pm_k=e^{i v^\pm_k}, \ \ \xi =e^{i \la}          . \ll{Lnul1} \ee
The validity of  (\re{cond}) by both
 (\re{llm-ul}) and (\re{Lnul1})
  gives  the explicit form of the braiding matrix  
$ Z^{(2)}_{12}=e^{i{ \al \ov 2} \sigma^3} \otimes I $, which differs
clearly from the previous case. Therefore, as established above,
 this discrete NSG  (\re{Lnul1}) must be a genuine    quantum
integrable NUL model, which  satisfies 
(\re{zlzl1u}) 
 and the braided YBE (\re{bqybel}).
For  finding  the full set
 of its mutually commuting 
conserved operators:  $ c_{\pm j}, j=1,2. \ldots, N$ we have to 
factorize the trace identity, which however
 becomes easy for the present $Z$ matrix 
yielding $\tilde \tau(\xi)=tr(e^{-i{ \al \ov 2} \sigma^3} T^{nul}(\xi) )$
 as
the generating function: $ \tilde \tau(\xi)=\sum^N_{\pm j}c_{\pm j} \ \xi^{\pm j}
$.
We may define the  Hamiltonian  of this new integrable  model
 as  $H= \ha((c_N)^{-1}c_{N-2} + (c_{-N})^{-1}
c_{-(N-2)})
= \sum^N_j \cos (v^+_j -v^-_j +{\al })$ and  name it 
   as  nonultralocal SG model, due to its
resemblance with the standard SG. 
Note that noncanonical commutation relations  (\re{cr12})
 involving different
sites, which must  be used
 for deriving the dynamical equations, induce nearest-neighbor interaction
 in the model. Notice also that by transforming 
$\xi \to {1 \ov \xi}, W^+_k \leftrightarrow 
 W^-_k $ in  (\re {Lnul1}) we would get   a dual though
 similar integrable NSG model.
\\
2.3. {\it Quantum light-cone sine-Gordon  model } 

Since  quantum   formulation through the braided YBE 
is not available for  this well known  nonultralocal classical
model, our first goal  is to discover such a formulation, 
for which we  follow
 our above strategy and   
look  for its UL source model
 as a particular realization
of the general UL model (\re{L}).
We observe that the choice of   central elements 
   as
$c^-_a=0,\  c^+_a= \Delta$ or its complementary set
$c^-_a=\Delta, \ c^+_a= 0,$ with $a=1,2$  leads again to the simple
 algebra (\re{tcalg}). Therefore we can use  realizations similar to
(\re{cgcft}):
\be
S_k^ {\mp}=e^{\pm i p_k},  \ S_k^3 =   u_k + \epsilon {1 \ov \al} p_k 
  \ll{lcsg0} \ee
 with $\epsilon = \mp  $ for the two complementary ({\it left} , {\it right})
cases  and generate  from (\re{L}) (after multiplying with $\sigma^1$)
 the  {\it left} ( {\it right} )   Lax operator
$L^{(\mp) }_k(\la) $ as 
\be
L^{(\mp) ul}_k{(\la)} = e^{ip_k\si ^ 3} + 
  \Delta \xi^{\pm 1}\left( e^{i(\pm \al u_k -  p_k)}\si ^+ + 
    e^{i(\mp \al u_k + p_k)}\si ^ - \right ), \ \xi =e^{i \la} ,
          \ll{ul-+} \ee
representing again RTC type integrable UL model.
Acting now, for example,
  on the {\it left}
 Lax operator $L^{(-) }_k(\la) $ of (\re{ul-+}) by a gauge matrix
$  D^{(3)}_j (\al)= 
e^{-i {\al } u_j \sigma ^3}\ \ $, which 
may be noticed to be simply
the {\it square} of that found in the previous cases:
$  D^{(3)}_j (\al)= (D^{(2)}_j)^2$, 
we 
 construct a  NUL Lax operator
\be
L^{(-)lcsg}_j{(\la)} =
   e^{i(p_j-\al \nabla u_j)\si ^3} + 
  \Delta \xi \left ( e^{-i(p_j+\al  u_{j+1})} \si ^+
    + e^{i(p_j+\al  u_{j+1})} \si ^- \right ) 
, \quad \nabla u_j \equiv  u_{j+1}- u_{j}. 
\ll{lcsg}\ee
We may check directly 
   that    
$L^{(-) ul}_k{(\la)}$ in (\re {ul-+}) as well as  $L^{(-)lcsg}_j{(\la)}$     
in (\re {lcsg}) respect   condition (\re{cond})  
with  the  braiding
matrix 
$ Z_{12}^{(-)(3)}=e^{i \al \sigma^3 \otimes \sigma^3} $ and
 therefore conclude that the NUL  Lax-operator
 (\re{lcsg})  together with   $R$-matrix (\re{R-mat}) and this
  $Z^{(-)(3)}$-matrix satisfies braiding  relation
(\re{zlzl1u}) and  the  BYBEs
 (\re{bqybel}), (\re{bqybet}).
 This   thus proves  the genuine quantum  integrability
of the associated model which, as we see below,  
 may be considered  as  
 the exact lattice version of the  quantum light-cone SG (LCSG) model.
  Therefore,  (\re {ul-+}) 
indeed is  an  UL source model
for the LCSG, which is  
  new as a quantum integrable NUL model.  
  At the field limit  ${\Delta \to 0}, $ when 
  we have $p_j \to \Delta \partial_t
u(x), \
  \al u_{j} \to u(x),\ \al u_{j+1} \to u(x) +\Delta  \partial_x
u(x) $, defining  $ \partial_t  
u \pm \partial_x
u =\ha  \partial_\pm 
u$,  it is not difficult to see that 
 (\re{lcsg}) reduces to
 $L^{(-)lcsg}_j{(\la)} \rightarrow  
  I +\Delta {\rm {U}}_-(x)$, where $ {\rm {U}}_-(x) 
=
{i \ov 2} \partial_- u(x) \si ^3 +    
   \xi ( e^{-i  u (x)} \si ^+ +  
     e^{i  u (x)} \si ^-)
$ yields  one of the  Lax pair :
 $\partial_- \Phi={\rm {U}}_- \Phi$ of the well known 
LCSG field model.

It is important to note  that the {\it right} operator 
$L^{(+)}_k{(\la)}$  in (\re {ul-+}) 
can be obtained formally from the {\it left}  
$L^{(-)}_k{(\la)}$  through  simple mapping ${ \xi} \to {1 \ov \xi},
\al \to - \al $ and  therefore the corresponding results 
for the complementary {\it right} LCSG can be recovered easily from 
that of the {\it left} one 
 through the same mapping.
Thus we  derive 
 the   NUL model  $L^{(+)lcsg}_j{(\la)} $
which  at the  field limit gives similarly 
  the other component
of the LCSG Lax pair: $ {\rm  {U}}_+(x) $.
  Zero curvature condition: \  $ 
\partial_- {\rm {U}}_+ -\partial_+ {\rm {U}}_+ + [ {\rm {U}}_+,
 {\rm {U}}_-]=0$ 
 involving both these  Lax operators yields finally
  the well known form of the sine-Gordon field 
equation in light-cone coordinates: $ \partial^2_{+-}u=2\sin2 u$. 

It is intriguing to remark  that, 
the UL source models found recently for the quantum left/right mKdV 
\c{gmkdv01} can be  given exactly by the same 
Lax operators (\re{ul-+}) discovered here for the   
light-cone SG model, though the   gauge operator related to the 
mKdV model coincides rather 
 with  $\sqrt  D^{(3)}_j (\al) $. 
\\
2.4. {\it Quantum mapping } \cite{Nijhof}

A rational quantum  mapping
related to the lattice KdV type equation 
may be described  by a   Lax operator 
\be U_n=L^{nul}_{2n}L^{nul}_{2n-1}, \ \ \mbox{where} \ 
L^{nul}_{aj}(\la _a)= \ha (1+\si^3_a) v_j +\si ^+ _a+ \la _a \si ^-_a , \
a=1,2 \ll{QM}\ee
which exhibits   NUL nature
due to the presence of discrete 
 current-like operator  $v_j \equiv v^+_j$ (\re{cr12}).
$L^{nul}_{j}$ and hence $U_n$ in (\re{QM})
 were shown  \cite{Nijhof} to  satisfy both  
braiding relation and  
BYBE  with the rational $R$-matrix 
$ R_{12}(\la_1-\la_2)= {\bf 1}+i{\al \ov 2} \frac {  P_{12}}{\la_1-\la_2},$
(which  may be obtained from 
  (\re{R-mat})
at $\al \to 0, \la \to 0$)
 and a spectral parameter dependent 
braiding matrix $Z^{(4)}_{21}(\la_2)
=
 {\bf 1} - \frac {i
 {\al } }{2 \la_2}
\si ^-\otimes  \si ^ + $  (similarly $Z_{12}(\la_1)$).

For finding the UL connection as asserted by 
the conjecture,
we look again for the source model
as a realization of the general construction  
and try to identify a suitable gauge matrix, as done in the above cases.
However since the present  model is associated with the rational $R$-matrix 
we have to start now from the ancestor Lax operator of
quantum integrable  rational UL models
 \c{kunprl},
which  is obtained as a  $\al \to 0$ (or $q \to 1$) limit of (\re{L}) in the form
 \be
L^{r-anc (ul)}{(\la)} = \left( \begin{array}{c}
 {c_1^0} (\la + {s^3})+ {c_1^1} \ \ \quad 
  s^-   \\
    \quad  
s^+    \quad \ \ 
c_2^0 (\la - {s^3})- {c_2^1}
          \end{array}   \right). \ll{LK} \ee
The basic operators  satisfy  the undeformed quadratic algebra
\be  [ s^+ , s^- ]
=  2m^+ s^3 +m^-,\ \ \ \ 
  ~ [s^3, s^\pm]  = \pm s^\pm 
  \ll{k-alg} \ee
with $m^+=c_1^0c_2^0, \  m^-= c_1^1c_2^0+c_1^0c_2^1$ and $c^n_a, a=1,2, \
n=0,1$ as
 central elements. For our construction
 we choose first  these elements as
 $c^0_1=i{\al \ov 2}, c^0_2= 0, c^1_1=-c^1_2=1$ to  derive the underlying algebra from 
(\re{k-alg}) in a simple form  
\be  [ s^+ , s^- ]
=  -i{\al \ov 2},\ \ \ \ 
  ~ [s^3, s^\pm]  = \pm s^\pm .
  \ll{kalg} \ee
We can therefore   realize this algebra simply as
\be  s^+ _k = -i \psi_k, \ s^- _k = \phi_k, \  s^3 _k = -i {2 \ov \al}
 \phi_k \psi _k \ll{qm} \ee
 through operators $ [\psi_j,\phi_k]=
 {\al \ov 2} \delta_{jk} $ to generate 
from (\re{LK}) (after multiplying from left by $\sigma^1$) an UL Lax operator
\be L^{ul}_j(\la _1)= \left( \begin{array}{c} -i \psi_j 
  \qquad \ \ 
   1  \\
    \quad  
    \la _1 -i \phi_j \psi _j \qquad \ \   
\phi_j
          \end{array}   \right) \ , \la_1= 1+ i{\al \ov 2} \la .
 \ll{map-ul} \ee
Remarkably this   Lax operator is associated with  a quantum integrable 
simple lattice NLS model \c{kunrag} and 
 satisfies  the YBE with rational $R$-matrix.
We find next   the gauge operator as  $D^{(4)}_j= I-\phi_j \si ^-
$, use a simple realization of the current-like operator satisfying
(\re{cr12}) as
$v^+_j=-i \psi_j + \phi_{j-1} $ 
 and perform the gauge transformation 
on  our UL model (\re{map-ul}) to   derive
 exactly  the Lax operator $L^{nul}_{j} $ of the 
 quantum mapping  (\re{QM}).
Note however that since we intend to recover the known form of $L^{nul}_{j} $
as given in \c{Nijhof}, we had to use  here a slightly different
form of gauge transformation: 
$D_jL^{ul}_j(\la _1)D_{j-1}^{-1}=L^{nul}_j(\la _1)$,
 compared to the rest of our  cases.  As a result 
 $L^{nul}_{j} $  would satisfy  NUL integrable relations, 
  which were  used in \c{Nijhof} and are complementary
 to those presented here and also in place of  
 (\re{cond}) a similar condition 
$D_{aj}L^{nul}_{bj}D_{aj}^{-1}=Z_{ab}^{-1}L^{nul}_{bj}$. 
 This establishes therefore  (\re{map-ul}) to be a  perfect gauge related
  UL source 
solution for  the well known NUL quantum mapping.
\\ \\ \noindent 3. {\bf Conclusion}

We conjecture an
ultralocal (UL) connection for every  nonultralocal (NUL)
quantum integrable model and establish the  yet unestablished 
such connections including some new ones.
We  find explicitly 
for  the well known quantum 
 NUL systems, e.g. CGCFT \c{babelon} and 
  quantum
mapping \c{Nijhof}, the corresponding 
gauge related   UL source models as the relativistic Toda chain (RTC)
 and a lattice NLS,
 which are well studied UL models. 
At the same time  we
 discover  two new quantum integrable NUL models, namely
a discrete NUL type  SG (NSG)
and an exact lattice light-cone SG (LCSG)  model
 and identify their
corresponding UL connections as  given through a Liouville and a RTC  
 model. 
Such an explicit UL  gauge relation of a NUL model would help in
extracting all necessary knowledge of a more complicated NUL system
including its  Bethe ansatz result through a more tractable and in most
cases 
already known    UL integrable model.  
 The NUL-UL connection
 should  become particularly useful, 
 when we are unable to tackle a  NUL integrable
 model   directly, 
 due to  nonfactorizability  of its trace identity or other reasons.
In such cases, thank to the present conjecture,
 we could  switch over to its gauge related UL model
  and use relation like (\re{monrel}): $\ti T^{nul}= T^{ul}$
for its UL description. The $\ti T^{nul}= L_{aN}^{nul} \cdots L_{a2}^{nul}
\ti L_{a1}^{nul}$  however would be an approximation,
 where one of its boundary 
terms is modified as $\ti L_{a1}^{nul}=D_1^{-1} L_{a1}^{nul}D_1 $.
It is remarkable that, though we have considered NUL models of completely
different nature with  no apparent relations between them, their gauge
related UL sources could be derived  
from the same ancestor model (or its $q
\to 1$ limit).
Such universality  of this
ancestor  is due to its general form, which is
 capable  of generating all possible quantum
integrable UL models within the specified class \c{kunprl}. Therefore we can
conjecture further that all other quantum integrable NUL models, not studied
here, also should have their gauge related UL  models derivable from the same
ancestor. Various reductions of this ancestor model, which  
generate different descendant UL models and serve as the source models
for the  NUL systems, are due to different choices of the central
elements leading to varied underlying algebras and hence various allowed
realizations. For generating the corresponding NUL models we need to find 
further the gauge operators, which depend in general on the structure
of  individual models as do also   the associated braiding matrices.
Among the four NUL models we considered, CGCFT and LCSG have the same underlying
algebra, while CGCFT and NSG have the coinciding gauge operator. The
braiding matrices of all these models however turn out to be different.

Since for
 our construction  we exploit the 
general algebraic scheme of  \c{kunprl}, we also unravel easily the  
important
 underlying algebraic structure for    each of the  NUL models. 
This  algebraic information along with their 
 gauge connections to 
UL models should make   quantum integrable NUL models 
more interesting and  easier objects 
to study,  boosting their
development which indeed deserves and needs more attention.

\end{document}